\documentclass[amsmath,amssymb]{revtex4}

\usepackage{color}

\usepackage{graphicx}
\usepackage{dcolumn}
\usepackage{bm}
\usepackage{epstopdf}
\usepackage{float}
\usepackage{soul}
\input epsf
\usepackage{subfigure}

\newcommand\beq{\begin{equation}}
\newcommand\eeq{\end{equation}}

\usepackage{color} 

\begin{document}
\title{Effect of viscosity on the dynamics of a non-equilibrium bubble in free-field and near a free-surface}
\author{Y. S. Kannan$^{\dagger}$}
\author{Saravanan Balusamy$^{\dagger}$}
\author{Badarinath Karri$^{\dagger}$}\email{badarinath@iith.ac.in}
\author{Kirti Chandra Sahu$^{\dagger\dagger}$}
\affiliation{
$^{\dagger}$Department of Mechanical and Aerospace Engineering, Indian Institute of Technology Hyderabad, Sangareddy 502 285, Telangana, India\\
$^{\dagger\dagger}$Department of Chemical Engineering, Indian Institute of Technology Hyderabad, Sangareddy 502 285, Telangana, India}

\begin{abstract}
The effect of viscosity on the behaviour of a non-equilibrium bubble is investigated experimentally, in two scenarios; firstly, when the bubble is generated in the bulk of the fluid (termed as ``free-field'' bubble) and secondly when the bubble is generated near a free-surface (termed as ``free-surface'' bubble). The bubble is created using a low-voltage spark circuit and its dynamics is captured using a high-speed camera with back-lit illumination. The viscosity of the surrounding fluid is varied by using different grades of silicone oil. For a ``free-field'' bubble, the bubble oscillates radially and as the viscosity of the liquid increases, the number of oscillations, as well as the time-period of each oscillation, are increased. At high viscosities, the bubble also becomes stable and does not disintegrate into smaller bubbles. For ``free-surface'' bubbles, two parameters, namely, the initial distance of the bubble from the free-surface and the viscosity of the surrounding fluid are varied. It is observed that beyond a certain initial distance of the bubble from the free-surface, the bubble behaves as a ``free-field'' bubble with negligible influence of the free-surface on its dynamics. This limiting initial distance decreases as the liquid viscosity is increased and is not dependent on the bubble radius. For these bubbles, different behaviours of the free-surface in each liquid are also presented as a function of the two parameters.

\end{abstract}

\maketitle
\section{Introduction}
\label{sec:intro}
Non-equilibrium bubbles have large pressure difference across their interface, which leads to radial (spherical) oscillations when the bubble is placed in a uniform pressure field, far from any nearby boundary. When the pressure difference across the interface is small, it leads to small amplitude oscillations (stable cavitation), where the contraction and expansion are symmetrical. When the pressure difference is large, the bubble expands and collapses rapidly and violently; such behaviour is also known as inertial cavitation.

The spherical oscillations of a non-equilibrium bubble were first analytically modelled by Rayleigh \cite{rayleigh1917viii} for liquids in the inviscid limit for an empty bubble. This was later extended to gas filled bubble by Lamb \cite{lamb1932} and the effect of viscosity was included in the model by Poritsky \cite{poritsky1952collapse}. Additional complexities of liquid compressibility and acoustic radiation from the bubble were considered in later analytical models, e.g., the Gilmore model \cite{strasberg1959onset} and the Keller-Miksis model \cite{crum1980measurements}. Apart from these analytical models, several other researchers (see e.g. Ref. \cite{plesset1977bubble} and references therein) have also studied the behaviour of such ``free-field'' bubbles both numerically and experimentally. There exist however only a few studies on the effect of viscosity on the dynamics of such free-field bubbles which are discussed in the next paragraph.

Jomni {\it et. al.} \cite{jomni2000experimental} used photodiode signals to characterise the dynamics of a ``free-field'' bubble generated using an electrical-pulse in viscous di-electric fluids. They reported that the expansion and collapse times of the bubble increased with an increase in the viscosity of the liquid. Englert {\it et al.} \cite{englert2011luminescence} also used photodiode signals to study the luminescence from laser generated bubbles as the viscosity of the surrounding liquid is varied. They reported that the number of bubble oscillations increased with an increase in the viscosity of the surrounding liquid; however, no direct visual evidence (through imaging) was provided about the bubble dynamics. Karri {\it et. al.} \cite{karri2011151} used high-speed imaging to study ``free-field'' bubble dynamics in viscous fluids using glycerol-water mixtures of viscosities 450 mPas and 1100 mPas. They presented the variation of the bubble radius versus time for these fluids and compared the results with that observed in water. It was, however, not a detailed study of the effect of viscosity on ``free-field'' bubble dynamics.

A bubble in the presence of a boundary (free-surface or solid wall) oscillates aspherically leading to the formation of a re-entrant jet during its collapse phase. In the presence of a solid boundary, the jet is directed towards the boundary \cite{plesset1971collapse,kornfeld1944destructive,naude1961mechanism,benjamin1966collapse} which can cause erosion and subsequent damage, and therefore has been an ongoing subject of research. For a bubble near a free-surface, numerical simulations \cite{blake1987transient,dommermuth1987numerical,wang1996strong,wang1996nonlinear} and experiments \cite{chahine1977interaction,blake1981growth,zhang2013experiments} have reported that the jet directs away from the free-surface. As a result of this jet, the bubble becomes toroidal and a reaction force is observed at the free-surface in the form of a spike and skirt. Chahine's experiments \cite{chahine1977interaction} showed that the re-entrant jet speed in bubbles near a free-surface are smaller than those observed for bubbles near a solid boundary. Robinson {\it et al.} \cite{robinson2001interaction} studied the dynamics of single and two bubbles near a free-surface both experimentally and numerically. Zhang {\it et al.} \cite{zhang2016experimental} have experimentally studied the bubble shape, centroid migration and jet-tip velocity for bubbles collapsing near a free-surface. In all of the above-mentioned studies, however, the effect of the surrounding fluid viscosity on a bubble near a free-surface has not been considered. 

In the present study, we provide a detailed experimental investigation of the effect of viscosity on the dynamics of free-field bubbles as well as bubbles near a free-surface. Liquids of different viscosity are obtained by using different grades of silicone oil as the surrounding fluid. For free-field bubbles the bubbles were created far away from any nearby boundaries. For free-surface bubbles, an additional parameter, the stand-off distance, $h_{ND}$, is varied in the experiments. Here, the ``stand-off'' distance, $h_{ND}$ is defined as the ratio of the initial distance of bubble center from the free-surface, $h_{f}$, to the maximum bubble radius, $R_{max}$, i.e. $h_{ND} \equiv {h_{f}/{R_{max}}}$. The variation of viscosity and stand-off distance leads to a change in the bubble re-entrant jet behaviour as well as liquid free-surface dynamics, which is systematically studied through high-speed imaging.

\section{Experimental set-up and procedure}
\label{sec:expt}

\begin{figure}[h]
\centering
\includegraphics[width=0.5\textwidth]{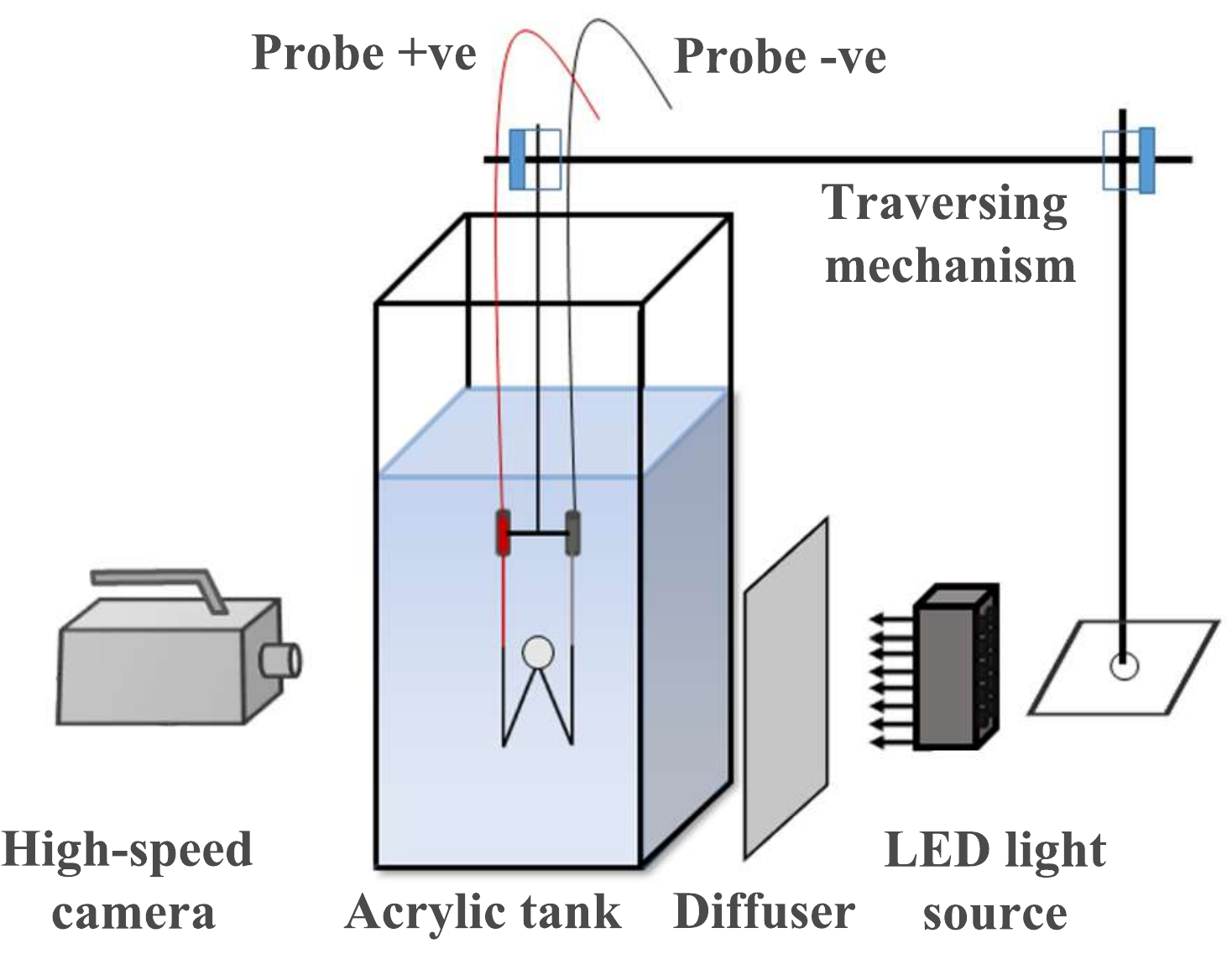}
\caption{Schematic diagram (not to scale) showing the experimental set-up.}
\label{fig1}
\end{figure}

The schematic of the experimental set-up is shown in Fig. \ref{fig1}. It consists of four parts: (i) an acrylic tank to hold the liquid, (ii) a low-voltage spark circuit to create the bubble, (iii) a traversing mechanism to position the electrodes and (iv) a high-speed camera with lighting system to record the bubble dynamics.

An acrylic tank of inner dimensions 400 mm $\times$ 400 mm $\times$ 500 mm is used for holding the surrounding liquid. For all the experiments, the tank was filled up to 350 mm of its height with the liquid. In order to study the effect of  the viscosity, we used five different liquids, namely, deionized water (kinematic viscosity, $\nu = 1$ cSt), and silicone oils of $\nu$ equal to 10 cSt, 100 cSt, 500 cSt and 1000 cSt. For experiments on ``free-field'' bubbles, it was ensured that the distance of the bubble from all adjacent boundaries is greater than 10 times the maximum radius of the bubble, $R_{max}$. In experiments involving bubbles near a free-surface, bubbles are created at different values of the ``stand-off'' distance $h_{ND}$ in each of the five liquids and the dynamics therein is recorded. 

The bubble is generated using a modified version of a low voltage spark circuit described by Goh {\it et al.} \cite{goh2013}. The circuit uses the principle of charging two capacitors (equivalent capacitance = 6900 $\mu$F) to a voltage up to 180 V through a direct-current (DC) power supply and then allowing them to be short-circuited/sparked through a pair of contacting electrodes. The electrodes are made of 0.1 mm diameter copper wires. During the experiments, they are placed inside the liquid and crossed so as to be in contact at a single point. As the electrodes are short-circuited by discharging the stored energy in the capacitors, a spark is produced, along with generation of a vapour bubble with the bubble centre at the point of contact of the electrodes. For a given voltage, the bubble size can be kept consistent by using electrode wires of a fixed length; e.g. for 60 V and 10 mm long electrode, the size of the bubble generated is $3.6 \pm 0.3$ mm. The leftover charge in the capacitor is drained out using a discharging circuit before the next experiment is performed. The charging, discharging and sparking circuits are connected to relays and a metal-oxide-semiconductor field-effect-transistor (MOSFET), and the different parts of the circuit are controlled using National Instruments-Data Acquisition Unit (NI-DAQ) (model USB-6008) through a LabVIEW program. In order to ensure consistency in experimental conditions we have taken the following precautions. The voltage of the spark circuit is kept constant for each set of experiments, namely 60 V for ``free-field'' bubbles and 100 V for ``free-surface'' bubbles. Experiments corresponding to Fig. 6 are conducted at higher voltages to achieve bigger bubble size. The length of each of the electrodes was kept constant at L = 10 mm for each experiment to ensure consistent bubble size \cite{goh2013}. A traversing mechanism is used to move the crossed electrodes (i.e the centre point of the bubble) within the liquid in all the directions in order to position the bubble within the liquid. All experiments presented are conducted at a temperature of $25^{o}$C and 1 atmospheric pressure.

A high-speed camera (Photron SA 1.1) is used to acquire 12 bit images of 256 $\times$ 320 pixels with a nominal resolution of 16.7 pixels/mm in case of experiments associated with the free-field bubble, and 12 bit images of 352 $\times$ 512 pixels with a nominal resolution of 5 pixels/mm in case of experiments of the free-surface bubble. The camera is equipped with a Sigma zoom lens of 28-300 mm focal length (f/3.5). The back-lit illumination is realized by using high-intensity LEDs (model 900445, Visual Instrumentation Corp.) with a luminous output of 12,000 lm. A tracing paper is used to diffuse the light uniformly. The focal plane of the camera is always at the initial center of the bubble. The images presented in the figures have been cropped to remove extraneous details. 

In order to capture the dynamics of the free-field bubble, images are acquired at 50,000 frames per second (fps) with an exposure time of 10 $\mu$s. In this case, the bubble is positioned close to the centre of the filled-up portion of the tank, and each experiment is repeated five times to check for the consistency of the bubble size. In case of free-surface bubble experiments, the images are acquired at 30,000 fps with an exposure time of 25 $\mu$s. In this case, the bubbles are created at different values of the stand-off distance $h_{ND}$ varying from 0.2 up to ${h_{lim}}$. Here, ${h_{lim}}$ refers to the stand-off distance at which the effect of the free-surface on the oscillating bubble becomes negligible and the bubble in essence behaves like a free-field bubble. For these experiments, the dynamics of both the bubble and the free-surface are captured in the same image for each liquid. 

\begin{figure}[H]
	\centering
	\includegraphics[width=0.75\textwidth]{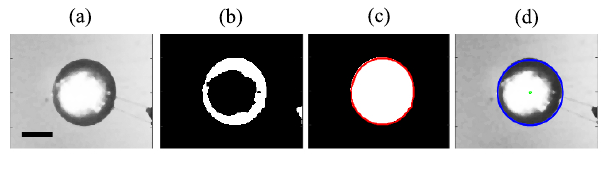}
	\caption{Series of events in the image processing using the in-house developed {MATLAB\textregistered} code. (a) Raw image, (b) background subtracted binarized image, (c) contour extraction (red colour circle) and (d) the least square fit (blue colour circle). The scale bar shown in the first frame corresponds to a length of 2 mm.}
	\label{fig9}
\end{figure}

\begin{figure}[H]
	\centering
	\includegraphics[width=0.6\textwidth]{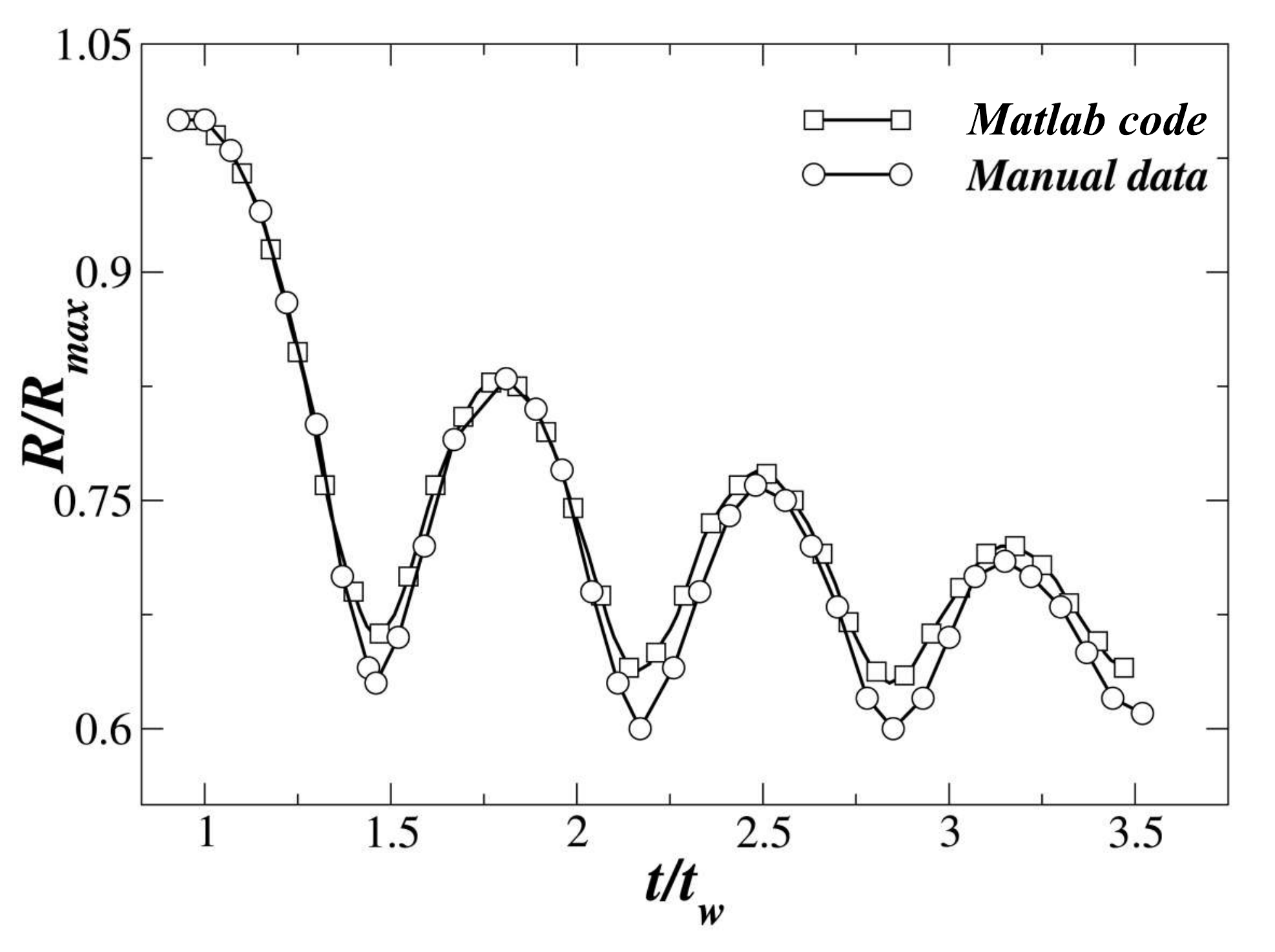}
	\caption{Comparison of the temporal evolutions of the dimensionless radius of a free-field bubble in 1000 cSt silicone oil obtained manually and using the {MATLAB\textregistered} code.}
	\label{fig9_1}
\end{figure}

The radius value of the bubble from each frame of the video is obtained from the in-house developed {MATLAB\textregistered} code. Initially, the code reads data from the raw image file (see Fig. \ref{fig9} (a)) and background subtraction of the image is done using the data of the first frame of the video. Then the image is binarized using the Otsu threshold \cite{Otsu1979} (see Fig. \ref{fig9} (b)). The hole within the bubble is filled to binarize the bubble completely and the contour (red colour circle) is extracted and is overlapped on the binarized image as shown in Fig. \ref{fig9} (c). The least square fit technique is applied to the obtained contour profile to know the radius and center of the bubble. The circle (blue colour circle) obtained from the least square fit is overlapped on the initial raw image in Fig. \ref{fig9} (d). A test experiment of a free-field bubble in 1000 cSt silicone oil is considered to validate the {MATLAB\textregistered} code. The radius evolution of the bubble from the test experiment is obtained manually by taking the average radius in the vertical and the horizontal directions. A comparison plot of the evolution of bubble non-dimensional radius versus non-dimensional time for the manual data and result obtained from the {MATLAB\textregistered}code is shown in Fig. \ref{fig9_1}. The error between results obtained manually and from the processing of the {MATLAB\textregistered}code is found to be less than 5 \%. It should be noted here that the Otsu threshold could be applied to our images only after the spark is subsided. Hence, the values of the droplet radius reported here do not start from the dimensionless time, $t/t_{w} = 0$.

\begin{figure}[H]
	\centering
	\includegraphics[width=0.75\textwidth]{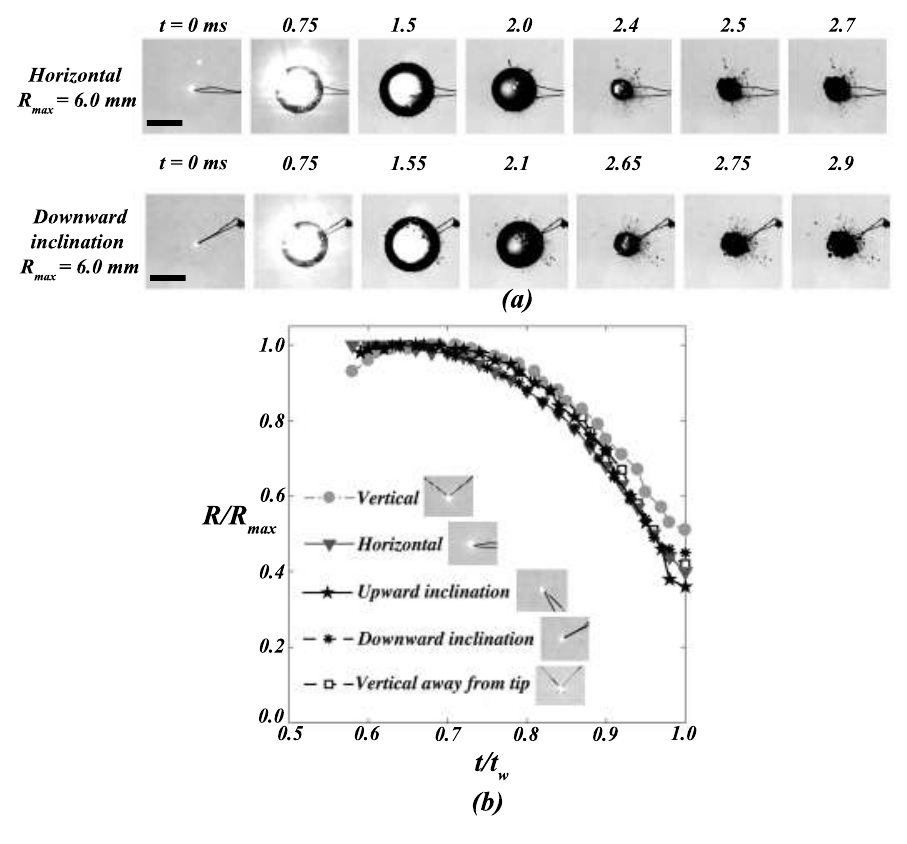}
	\caption{The dynamics of a free-field bubble in deionized water for different electrode orientations. (a) The time series images of the evolution of bubble in two different electrode orientations, namely, horizontal (first row) and downward inclined (second row). Time is shown on top of each image. The scale bar shown in the first image of each row corresponds to a length of 10 mm. (b) The variations of the dimensionless radius with the dimensionless time for bubble in different orientations.}
	\label{fig2}
\end{figure}

It should be noted here that the contacting electrodes may have different orientations when placed together (please see Fig. \ref{fig2} (b) for example) and the contact point of the electrodes may not be exactly at the tips (leading to a slight variation in the electrode length ``L''). A question that could arise is if these factors can affect the bubble dynamics. In order to address this question comprehensively, we examined the effect of (a) the position of electrode contact point at the electrode tip and (b) the orientation of the electrodes on the spark generated bubble. Free-field bubble experiments were conducted in deionized water for five different orientations of the electrodes, namely (i) vertical (electrodes are positioned from top), (ii) horizontal (electrodes are positioned from side), (iii) upward inclined electrodes, (iv) downward inclined electrodes and (v) vertical with electrodes contact point away from the tips of the electrodes. Fig \ref{fig2} (a) presents the sequence of images showing the time series events of the dynamics of bubble for the horizontal orientation (first row) and downward inclined electrodes (second row). For both the orientations the time evolution of different phases of the bubbles are similar. For example the maximum size of the bubble happens at 1.5 ms for horizontal orientation and at 1.55 ms for downward inclined electrodes and the minimum happens at 2.4 ms for horizontal orientation and at 2.65 ms for downward inclined electrodes. For all the five orientations the bubble evolution is shown in the form of a non-dimensional plot of $R/R_{max}$ versus $t/t_{osc}$ in Fig. \ref{fig2} (b). Here, $t_{osc}$ is the time period of the first oscillation of the bubble in the corresponding experiment. From both Figs. \ref{fig2} (a) and \ref{fig2} (b) we can observe that slight changes in the orientation of the electrode or the contact point at the electrode tip has only a minimal effect on the bubble dynamics.

\pagebreak

\section{Results and discussion}
\label{sec:dis}

\subsection{Effect of viscosity on ``free-field'' bubbles}
\label{subsec:freefield} 

Table \ref{T1} shows the averaged maximum bubble radius ($R_{max}$) over five repeated experiments for each fluid, the standard deviation of $R_{max}$ (expressed as a percentage of $R_{max}$), the Reynolds number and the Weber number of a free-field bubble for all the five surrounding liquids. The fluid properties of the surrounding liquids, such as the values of density ($\rho_L$), and surface tension ($\sigma$) are also listed for completeness.

\begin{table}[h]
	\begin{center}
		\caption{Fluid properties and dimensionless numbers associated with ``free-field'' bubbles in various fluids.} \label{T1} 
		\begin{tabular}{|c | c | c | c | c | c | c | c | c |}	
			\hline
			S.No & Working & Density, $\rho_L$ & Surface tension ($\sigma$)  & $Re$ & $We$ & $R_{max}$ & $R_{max}$-Standard \\ 
			& fluid & (kg m$^{-3}$) & (mN m$^{-1}$) &  &  & (mm) & deviation (\%) \\ 
			\hline
			1 & Deionized water  & 997 & 71.99 &  17736 & 1197 & 3.6  & 3.5 \\ 		\hline
			2 & 10 cSt silicone oil  & 930 & 20.1 & 1431 & 2692 & 3.6 & 5.9 \\
			\hline
			3 & 100 cSt silicone oil  & 960 & 20.9 &  98 & 1322 & 3.3 & 4.8 \\
			\hline
			4 & 500 cSt silicone oil  & 970 & 21.1 &  17 & 995 & 3.2 & 3.2 \\
			\hline
			5 & 1000 cSt silicone oil  & 970 & 21.1 & 7 & 637 & 3.3 & 1.2 \\
			\hline
			
		\end{tabular}
	\end{center}
\end{table}

The Reynolds number (Re) and Weber number (We) are defined as 
\begin{equation}
Re = { V R_{max}\over\nu_L},
\label{eq1}
\end{equation}
\begin{equation}
We = { \rho_L V^2 R_{max}\over \sigma},
\label{eq3}
\end{equation}
where, $\nu_L$ is kinematic viscosity of the fluid and $V$ is the average velocity of the bubble wall given by
\begin{equation}
V = {R_{max}-R_{min} \over {t_{c}}}.
\label{eq2}
\end{equation}

Here, $R_{min}$ represents the minimum observed radius of the bubble and $t_{c}$ represents the experimental collapse time, which is the time taken by the bubble wall from first $R_{max}$ to first $R_{min}$. 

Weber number values range from 1197 for deionized water to a value of 637 for 1000 cSt silicone oil indicating that inertia is dominant as compared to surface tension. The comparison of Reynolds number and Weber number for all the five fluids indicates that for 500 cSt and 1000 cSt silicone oils viscosity is dominant. 

It can also be noted from Table \ref{T1} that the maximum standard deviation of $R_{max}$ is 5.9 \%, which is similar to the value reported by Goh {\it et al.} \cite{goh2013}. The values of the maximum bubble size ($R_{max}$) are also observed to be similar for all the fluids used in the present experiments when all other parameters are kept constant.

\begin{figure}[H]
	\centering
	(a) \hspace{1.3cm} (b) \hspace{1.3cm} (c) \hspace{1.3cm} (d) \hspace{1.3cm} (e) \\
	\includegraphics[width=0.56\textwidth]{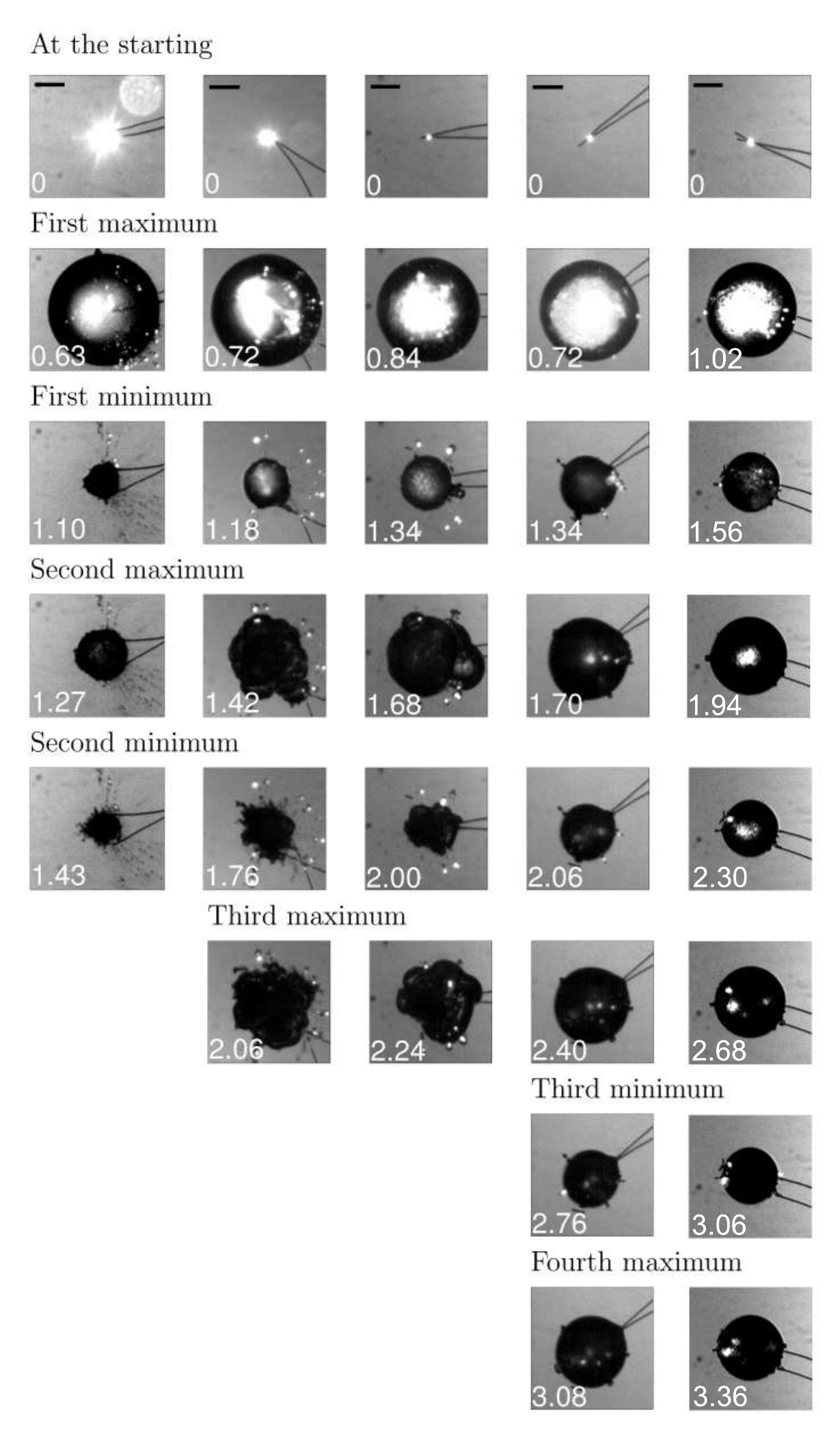}
	\caption{The dynamics of a free-field bubble in liquids of different viscosities: (a) deionized water,  (b) 10 cSt silicone oil, (c) 100 cSt silicone oil, (d) 500 cSt silicone oil and (e) 1000 cSt silicone oil. The bubbles in panels (d) and (e) exhibit further oscillations, which are not shown here. The scale bar shown in the first frame of each panel corresponds to a length of 2 mm. The number at the bottom left corner of each frame in the panel is time, $t$ in ms; $t = 0$ corresponds to spark initiation at which the bubble is incepted. The bright circular spot at the top right corner of the first frame in panel (a) is a reflection of the spark from the wall of the tank. The bubbles are shown at the maximum and minimum sizes in each oscillation as observed in the recorded videos.}
	\label{fig3}
\end{figure}

\begin{figure}[H]
	\centering
	\includegraphics[width=0.56\textwidth]{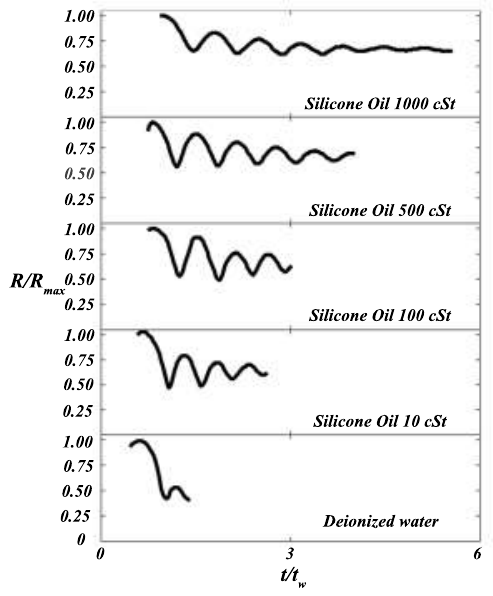}
	\caption{The variation of dimensionless radius of the bubbles $(R/R_{max})$ versus dimensionless time $(t/t_{w})$ for different surrounding liquids. The uncertainty in measuring $(R/R_{max})$ is 0.01.}
	\label{fig4}
\end{figure}

Fig. \ref{fig3} shows the dynamics of a free-field bubble captured inside the five liquids considered in the present study. Each column of images corresponds to one particular liquid. In every column, the maximum and minimum sizes of the bubble for each oscillation are shown, and the oscillation number is indicated above each row. The number in each frame indicates time in millisecond (ms), where $t=0$ corresponds to the initiation of the spark. Note that the `minimum' size of the bubble in each oscillation, as presented in Fig. \ref{fig3}, may not represent the actual minimum for that oscillation. This is because obtaining the true minimum size would need a frame rate of about 10 million fps to have enough temporal resolution, which is beyond the capability of the high-speed camera used in our experiments. The $R_{max}$ of the bubble shown, however, is fairly accurate as the bubble stays at a value close to $R_{max}$ for a considerable amount of time as can be seen from Fig. \ref{fig4} i.e. $dR/dt$ at $R_{max}$ is very small compared to $dR/dt$ at $R_{min}$.

Fig. \ref{fig3}(a) shows an oscillating bubble in deionized water. In this case, the bubble collapses very rapidly, and only two clear oscillations could be observed. The bubble expands to its maximum radius $R_{max}$ at $t= 0.63$ ms, followed by a contraction and the first minimum of its oscillation at $t=1.1$ ms. After this, the bubble undergoes small amplitude rapid oscillations, leading to a second maximum at $t=1.27$ ms. The bubble disintegrates in the surrounding fluid after $t=1.43$ ms, beyond which further maxima and minima cannot be distinguished clearly.

Figs. \ref{fig3}(b) and (c) show the oscillations of the non-equilibrium bubble in 10 cSt and 100 cSt silicone oils, respectively. In both these liquids, three distinct oscillations can be observed before the bubble disintegrates into the surrounding fluid after the third maximum is observed. The bubble completes the first oscillation at $t=1.18$ ms (10 cSt) and $t=1.34$ ms (100 cSt), which are greater than the corresponding values observed in case of deionized water (Fig. \ref{fig3}(a)). The second minimum is observed at $t=1.76$ ms (10 cSt) and $t=2$ ms (100 cSt), which are again greater than that observed in Fig. \ref{fig3}(a). A close observation of Figs. \ref{fig3}(b) and (c) also indicates an increase in the number and time period of each oscillation as the viscosity is increased. This is further corroborated by the observations for 500 cSt silicone oil (Fig.  \ref{fig3}(d)) and 1000 cSt silicone oil (Fig.  \ref{fig3}(e)). In Figs. \ref{fig3}(d) and (e), we actually see more than four distinct oscillations; however, only four oscillations are shown due to space. Inspection of Fig. \ref{fig3}, in a row-wise manner, also confirms that for the corresponding stages in oscillations (appearance of maxima and minima), the time period increases as the viscosity of the surrounding fluid is increased. At higher viscosities, the bubbles also appear to be more stable in terms of maintaining a spherical shape over a longer period of time without disintegrating into the surrounding fluid.

To compare the oscillations of the bubble for all five liquids, the variation of the dimensionless radius of the bubble, $R/R_{max}$ is plotted with respect to dimensionless time, $t/t_{w}$ in Fig. \ref{fig4}. Here, $t_{w}$ is the time period of first oscillation of the free-field bubble in deionized water. We considered $t_{w}$ for non-dimensionalising time so as to clearly observe the change in the time period of oscillation as viscosity of the fluid is increased. As viscosity varies from 1 cSt to 1000 cSt, the $t/t_{w}$ values for the first oscillation increases from 1 to 1.46. In other words, the time period of first oscillation increases with an increase in the viscosity of the surrounding fluid. The number of clear oscillations also increases with an increase in viscosity, as also discussed in Fig. \ref{fig3}. 

In order to explain the dynamics observed in Figs. 3 and 4, let us first consider the evolution of a spark induced ``free-field'' bubble in any given fluid. The initiation of the spark leads to a bubble with a large internal pressure as compared to the surroundings because of the vapor and gas due to parts of the electrode burnt by sparking/short-circuiting. As the bubble expands driven by this pressure difference (internal to external), it expands till the internal pressure becomes equal to the surrounding pressure and then continues to expand because of the inertia of the bubble wall. The bubble thus over-expands leading to a lower internal pressure compared to the surroundings and then it starts to collapse again reaching equilibrium (in pressure) but shrinking further due to inertia. This cycle would continue infinitely if there were no opposing forces. However, there is an opposing force to this inertia-driven dynamics due to the viscosity of the surrounding fluid. The movement of the bubble wall is opposed by the viscous forces due to the surrounding fluid i.e. while the bubble is expanding the viscous force tries to restrict the expansion of the bubble and in the collapse phase of the bubble the viscous force tries to restrict the collapse of the bubble. Thus an increase in viscosity leads to a slower expansion and also a slower collapse due to the larger resistance offered by the fluid. This probably explains why we observe an increase in the time period of the oscillations of a free-field bubble. The maximum bubble size on the other hand is determined by the input energy which is similar across all the fluids because of the input voltage and the electrode length being the same in all the experiments. Thus we do not see much of a difference amongst the maximum bubble sizes across fluids of different viscosities (see Table \ref{T1} with a difference of only about 8.1 \%.

It is relevant to note here that the observations presented above for ``free-field'' bubbles in viscous fluids agree well with earlier studies \cite{jomni2000experimental, englert2011luminescence, karri2011151} (see also Sec. \ref{sec:intro}). In addition to providing more details on bubble dynamics, our study differs from Refs. \cite{jomni2000experimental,englert2011luminescence} in terms of the experimental approach used and the length scales involved. The conclusions in both studies \cite{jomni2000experimental,englert2011luminescence} were made using the signals obtained from the photodiode (not direct visualization) and the bubbles considered therein were of the order of a few micrometers. The study by Karri {\it et al.} \cite{karri2011151} is more similar to the present study in terms of the approach, length scales and the observations. However, ``free-field'' bubbles was not the focus of their study and therefore they did not investigate the bubble dynamics to the detail that is presented here.


\subsection{Effect of viscosity on ``free-surface'' bubbles}
\label{subsec:freesurface} 

In this section, the dynamics of a bubble and the nearby free-surface are presented for ``free-surface'' bubbles as the viscosity of the surrounding fluid and the stand-off distance, $h_{ND} \equiv {h_{f}/{R_{max}}}$, are varied. Recall here that, $h_{f}$ is the initial distance of bubble center from the free-surface and $R_{max}$ is the maximum bubble radius.

A ``free-surface'' bubble experiences similar forces (inertial and viscous) driving its expansion and collapse as that of a ``free-field'' bubble but has an additional asymmetry since its top half is surrounded by the liquid ``free surface'' while the bottom half is surrounded by the fluid. This asymmetry in the bubble surroundings leads to a difference in the resistance offered by the viscosity of the fluid at the top and bottom half of the bubble. The ``re-entrant'' jet away from the free surface which is seen in the bubble during its collapse phase is because of this asymmetry. As the bubble expands, the top half expands faster compared to the bottom half and when the bubble starts to collapse, the top half also collapses faster compared to the bottom half due to lesser resistance by the fluid at the top half compared to the bottom half. The asymmetry in the collapse phase leads to the top portion collapsing into the bottom portion (which is still halfway through the collapse) which is observed as a ``re-entrant'' jet away from the free surface. 

We can now deduce that what might happen as the viscosity of the surrounding fluid is increased for a ``free-surface'' bubble. For similar $h_{ND}$ values, a fluid with higher viscosity will provide a greater resistance to both the bubble expansion and collapse. Specifically, for similar $h_{ND}$ values, we would expect that there will be a greater opposition to the ``re-entrant'' jet formation as the viscosity increases. The experimental results described below corroborate this expectation. We observed that for similar $h_{ND}$ values, the effect of an increase in viscosity is seen first as a slowing down of the formation of the ``re-entrant'' jet and subsequently a complete suppression of the jet at further increase in viscosity. 

Another possible effect of increase in viscosity on a ``free-surface'' bubble is on the expected variation in the limiting $h_{ND}$ value or $h_{lim}$. For any particular fluid as $h_{ND}$ is increased, one would expect that beyond a certain value of $h_{ND}$ the bubble is no longer influenced by the free-surface, i.e. it behaves like a ``free-field'' bubble. We can thus define a limiting $h_{ND}$ value ($h_{lim}$) as the minimum $h_{ND}$ at which the ``re-entrant'' jet is completely suppressed. Since a more viscous fluid offers greater resistance to the jet formation, one would expect the $h_{lim}$ value to be smaller for a more viscous fluid as compared to a fluid of lower viscosity. We also decided to test this assertion through experiments to determine the $h_{lim}$. For each fluid, in order to obtain the value of $h_{lim}$ for that fluid, we conducted experiments by increasing the $h_{ND}$ value till the ``re-entrant'' jet is completely suppressed. At this value of $h_{ND}$, we further varied the $h_{ND}$ at smaller increments and thus deduced the value of $h_{lim}$ with an uncertainty of about $\pm$ 0.1.

\begin{figure}[H]
	\centering
	\includegraphics[width=1.0\textwidth]{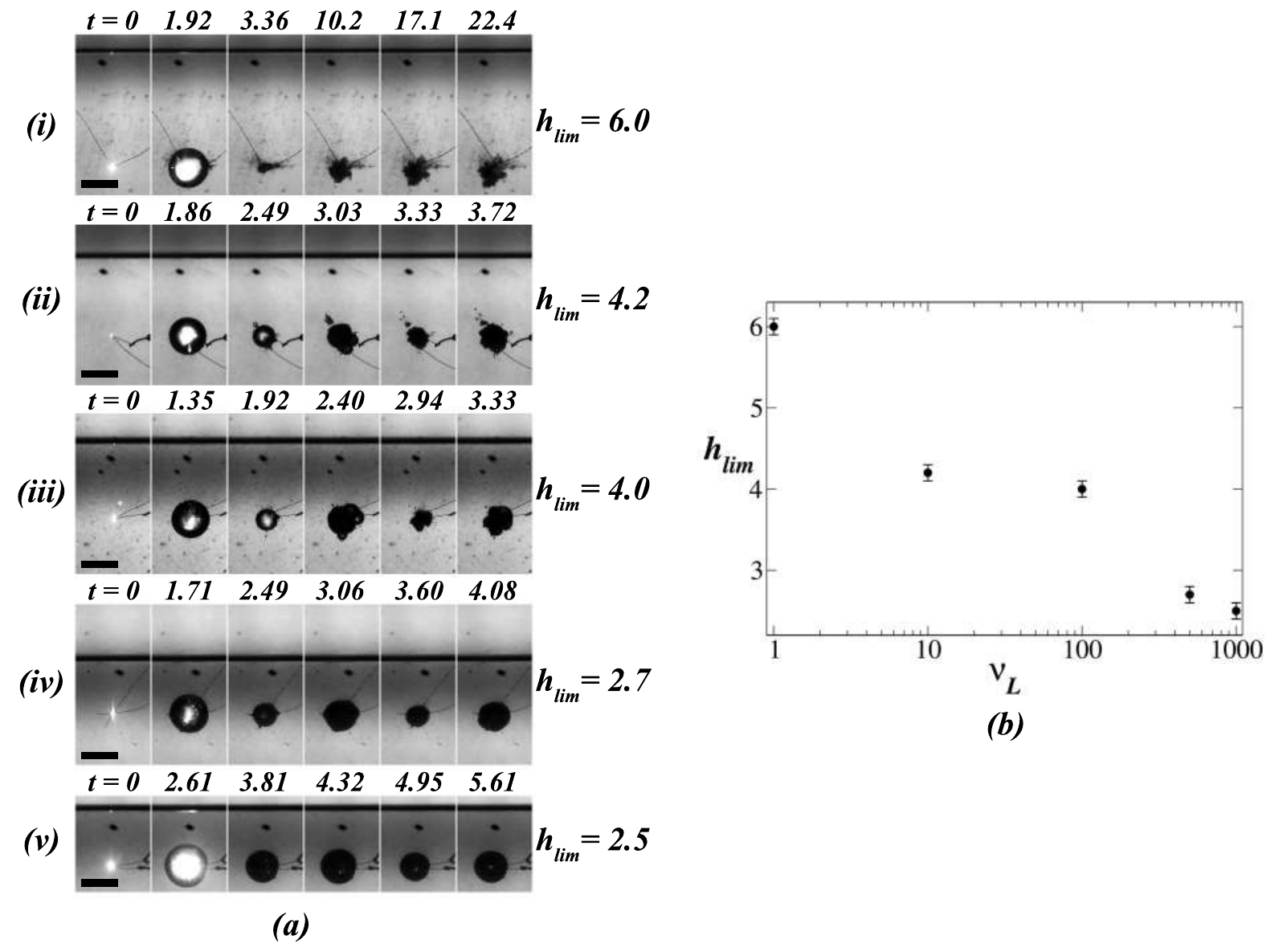}
	\caption{(a) The value of $h_{{lim}}$ and the dynamics of a bubble at $h_{{lim}}$ in different liquids. (i) Deionized water, $R_{max} = 5.8$ mm, (ii) 10 cSt silicone oil, $R_{max} = 5.6$ mm, (iii) 100 cSt silicone oil, $R_{max} = 5.6$ mm, (iv) 500 cSt silicone oil, $R_{max} = 5.6$ mm and (v) 1000 cSt silicone oil, $R_{max} = 6.6$ mm. The scale bar shown in the first frame of each row corresponds to a length of 10 mm. The number on the top of each image represents time in ms; $t = 0$ corresponds to spark initiation at which the bubble is incepted. Note that the black spot near the free-surface of liquid in all the panels is a particle on the sensor of the high-speed camera. (b) Plot of limiting stand-off distance ($h_{{lim}}$) versus kinematic viscosity ($\nu_{L}$) of the surrounding liquid.}
	\label{fig5}
\end{figure}

The values of $h_{{lim}}$ and the dynamics of the bubble at $h_{{lim}}$ in different liquids are presented in Figs. \ref{fig5}(i) (a) to (e). In these images the bubble does not form any re-entrant jet and the expansion and collapse of the bubble is nearly spherical. The values of $h_{{lim}}$ for deionized water and silicone oils of kinematic viscosity 10 cSt, 100 cSt, 500 cSt and 1000 cSt are obtained experimentally as described in the previous paragraph to be 6.0, 4.2, 4.0, 2.7 and 2.5, respectively. It can be clearly seen that the value of $h_{{lim}}$ decreases with an increase in the viscosity of the surrounding fluid.

\begin{figure}[H]
	\centering
	\includegraphics[width=1.0\textwidth]{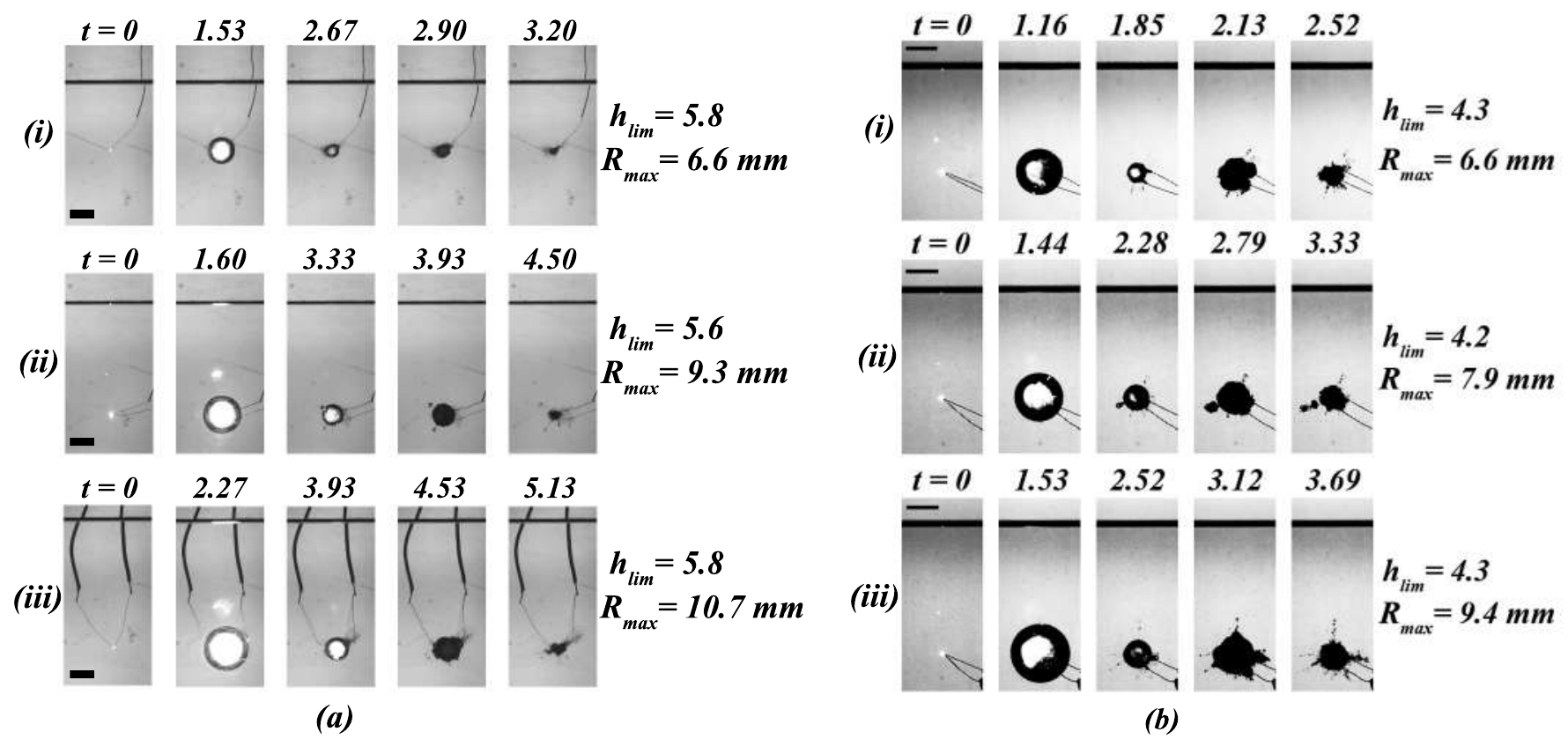}
	\caption{Experiments corresponding to $h_{{lim}}$ for (a) deionized water and (b) silicone oil of kinematic viscosity 10 cSt at three different values of $R_{max}$. The scale bar shown in the first frame of each row corresponds to a length of 10 mm. The number on the top of each image represents time in ms; $t = 0$ corresponds to spark initiation at which the bubble is incepted.}
	\label{fig6}
\end{figure}

A question that may arise is if the value of $h_{{lim}}$ for a particular fluid changes if the size of the bubble, i.e. $R_{max}$ is changed. We, therefore, considered deionized water and silicone oil of kinematic viscosity 10 cSt as the surrounding liquid and conducted additional experiments with three different values of $R_{max}$ (by varying the input voltage in the spark-generating circuit). Figures \ref{fig6}(i) and \ref{fig6}(ii) represent the effect of varying $R_{max}$ on the value of $h_{{lim}}$ for deionized water and silicone oil of 10 cSt kinematic viscosity respectively. From the four experimental observations for deionized water (Fig. \ref{fig5}(a) (i) and Fig. \ref{fig6}(a)) and silicone oil of 10 cSt kinematic viscosity (Fig. \ref{fig5}(a) (ii) and Fig. \ref{fig6}(b)), we observe that the limiting stand-off distance ($h_{{lim}}$) is nearly constant with a variation of only about 7 \% for deionized water and 2 \% for silicone oil. We can, therefore, conclude that the value of $h_{{lim}}$ is not dependent on the size of the bubble and depends only on the liquid used.

\subsection{Bubble and free-surface behaviour for $h_{{ND}}$ $<$ $h_{{lim}}$}

In addition to obtaining the value of  $h_{lim}$ for each of the fluids, we also conducted experiments over a number of $h_{ND}$ values ranging from 0.2 to $h_{{lim}}$ in each fluid. Our experiments reveal a variety of complex liquid free-surface behaviours for all the fluids as the bubble expands and collapses.

From all the ``free-surface'' experiments (nearly 200) the behaviour of fluid free-surface in each fluid for different $h_{{ND}}$ values is summarized in Table \ref{T2}. The observed behaviours of the fluid free-surface for these bubbles can be broadly classified as ``Spraying liquid sheet''(see Supplementary video 1), ``Ruptured free-surface with a jet on top'' (see Supplementary video 2), ``Unstable spike and unstable skirt'' (see Supplementary videos 3 and 4), ``Thin spike'' (see Supplementary video 5), ``Stable spike and stable skirt'' (see Supplementary video 6), ``Spike and a faster secondary jet'' (see Supplementary video 7) and ``Small rise of bulk fluid'' (see Supplementary video 8). Fig. \ref{fig7} shows the snapshots of these different dynamics which captures the distinct features of each of these behaviours. The features of two of these dynamics is described in detail in Fig. \ref{fig8}.

\begin{figure}[H]
	\centering
	\includegraphics[width=0.55\textwidth]{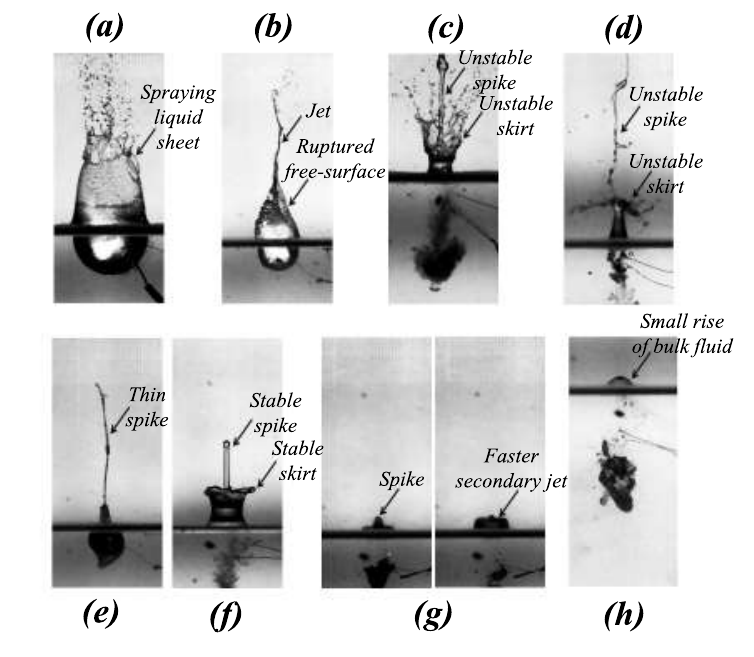}
	\caption{Images from the ``free-surface'' bubble experiments to describe different types of fluid free-surface behaviours. The thick black line separating the two fluids (Bottom - Working fluid, Top - Air) is the free-surface of the working fluid. The thin black lines in the bottom fluid are the electrodes. (a) Spraying liquid sheet in deionized water at $h_{ND} = 0.2$ (supplementary video 1) (b) ruptured free-surface with jet on top in 500 cSt silicone oil at $h_{ND} = 0.2$ (supplementary video 2) (c) unstable spike and unstable skirt in deionized water at $h_{ND} = 0.5$ (supplementary video 3) (d) unstable spike and Unstable skirt in 10 cSt silicone oil at $h_{ND} = 0.4$ (supplementary video 4) (e) thin spike in 500 cSt silicone oil at $h_{ND} = 0.3$ (supplementary video 5) (f) stable spike and stable skirt in deionized water at $h_{ND} = 0.8$ (supplementary video 6) (g) spike and faster secondary jet in 100 cSt silicone oil at $h_{ND} = 0.8$ (supplementary video 7) and (h) small rise of bulk fluid in 10 cSt silicone oil at $h_{ND} = 1.6$ (supplementary video 8).}
	\label{fig7}
\end{figure}



\begin{table}[H]
	\begin{center}
		\caption{Summary of the fluid free-surface behaviours observed for ``free-surface'' bubbles in various fluids.} \label{T2}
		\begin{tabular}{|c | c | c | c |}	
			\hline
			&  &   & Free-surface \\ 
			S.No & Working fluid  & $h_{{ND}}$ range & behaviour \\ 
			\hline
			& & 0.2 - 0.4 & Spraying liquid sheet. Fig. \ref{fig7} (a) \\ 	
			&   & 0.5 - 0.7 & Unstable spike and unstable skirt. Fig. \ref{fig7} (c) \\ 	
			1 & Deionized water & 0.8 - 0.9 & Stable spike and stable skirt. Fig. \ref{fig7} (f) \\ 
			&   & 1.0 - 1.5 & Spike and faster secondary jet. Fig. \ref{fig7} (g) \\ 
			&   & 1.6 - 1.9 & Small rise of bulk fluid. Fig. \ref{fig7} (h) \\ 
			\hline
			&   & 0.2 & Spraying liquid sheet. Fig. \ref{fig7} (a) \\ 	
			&   & 0.3 - 0.5 & Unstable spike and unstable skirt. Fig. \ref{fig7} (d) \\ 	
			2 & 10 cSt silicone oil & 0.6 - 0.8 & Stable spike and stable skirt. Fig. \ref{fig7} (f) \\ 
			&   & 0.9 - 1.2 & Spike and faster secondary jet. Fig. \ref{fig7} (g) \\ 
			&   & 1.3 - 1.6 & Small rise of bulk fluid. Fig. \ref{fig7} (h) \\ 
			\hline
			&   & 0.2 & Spraying liquid sheet. Fig. \ref{fig7} (a) \\ 	
			3 & 100 cSt silicone oil  & 0.3 - 0.6 & Unstable spike and unstable skirt. Fig. \ref{fig7} (d) \\ 	
			&   & 0.7 - 0.9 & Spike and faster secondary jet. Fig. \ref{fig7} (g) \\ 
			&   & 1.0 - 1.3 & Small rise of bulk fluid. Fig. \ref{fig7} (h) \\ 
			\hline
			&   & 0.2 & Ruptured free-surface with top jet. Fig. \ref{fig7} (b) \\ 	
			4 & 500 cSt silicone oil  & 0.3 & Thin spike. Fig. \ref{fig7} (e) \\ 	
			&   & 0.4 - 1.0 & Small rise of bulk fluid. Fig. \ref{fig7} (h) \\ 
			\hline
			&   & 0.2 & Ruptured free-surface with top jet. Fig. \ref{fig7} (b) \\ 	
			5 & 1000 cSt silicone oil  & 0.3 & Thin spike. Fig. \ref{fig7} (e) \\ 	
			&   & 0.4 - 0.9 & Small rise of bulk fluid. Fig. \ref{fig7} (h) \\ 
			\hline
			
		\end{tabular}
	\end{center}
\end{table}

\begin{figure}[H]
	\centering
	\includegraphics[width=0.55\textwidth]{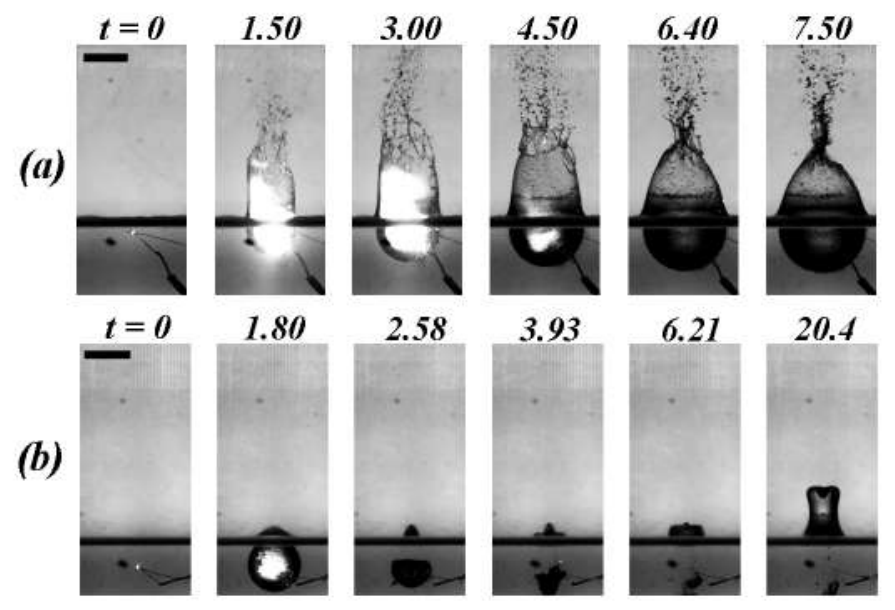}
	\caption{Experiments depicting the time series events of the dynamics of fluid free-surface for ``free-surface'' bubbles in two different fluids. (a) Free-surface dynamics of deionized water at $h_{ND}$ = 0.2 (Supplementary video 1). (b) Free-surface dynamics of 100 cSt silicone oil at $h_{ND}$ = 0.8 (Supplementary video 7). The scale bar shown in the first frame of each row corresponds to a length of 10 mm. The number on the top of each image represents time in ms; $t = 0$ corresponds to spark initiation at which the bubble is incepted. Note that the black spot near the free-surface of liquid in all the panels is some particle on the sensor of high-speed camera. }
	\label{fig8}
\end{figure}

Fig. \ref{fig8} (a) presents the time series events of the ``spraying liquid sheet'' dynamics from the ``free-surface'' bubble experiments in deionized water at $h_{ND}$ = 0.2. As the initial bubble center is very close to the liquid free-surface, the bubble expands into it breaking the free-surface ($t = 1.50 $ ms) leading to the formation of a spraying liquid sheet at $t = 3.00 $ ms. The time series events of the ``spike and faster secondary jet'' dynamics of the 100 cSt silicone oil at $h_{ND}$ = 0.8 is presented in Fig. \ref{fig8} (b). Due to the expansion of the bubble, a spike is observed at $t = 1.80 $ ms. As the bubble collapses at $t = 3.93 $ ms, a secondary jet is observed. At $t = 6.21 $ ms the secondary jet overtakes the spike and hence the secondary jet is termed as faster secondary jet. The dynamics mentioned in Fig. \ref{fig7} are provided as Supplementary videos 1 - 8.

A further note on Table \ref{T2} is pertinent at this stage. It can be seen by careful observation that the jet and spray/sheet that are observed on top of the free-surface become more stable as the viscosity is increased. For example, the``spraying liquid sheet'' and the ``unstable spike and unstable skirt'' which is observed for water, 10 cSt and 100 cSt silicone oils is no longer observed for 500 cSt and 1000 cSt silicone oils. Instead, we observe a ``ruptured free-surface with a jet on top''. Another observation is on the ``small rise of bulk fluid'' which is an indicator of the reduced effect on the free-surface by the bubble. For water, 10 cSt and 100 cSt, this is observed for $h_{ND} > 1$ while for 500 cSt and 1000 cSt it is observed at $h_{ND} < 1$. We can infer that higher viscosity liquids (500 cSt and 1000 cSt silicone oils) appear to have a different set of free-surface behaviors as compared to the lower viscosity liquids.

\section{Concluding remarks}
\label{sec:conc}
An experimental study has been conducted using high-speed imaging to observe the dynamics of a non-equilibrium bubble and the liquid free-surface for five liquids of different kinematic viscosities, i.e. deionized water, 10 cSt, 100 cSt, 500 cSt and 1000 cSt silicone oils. For each fluid, two types of bubbles were studied; firstly, a ``free-field'' bubble which is created far from any boundary and secondly a ``free-surface'' bubble which is created near a liquid free-surface at different distances from the liquid free-surface. In case of a free-field bubble, the time-period of oscillations of the bubble as well as the number of oscillations observed before the bubble disintegration are found to increase with an increase in the viscosity of the liquid. The bubble also becomes more stable as noted from the decrease in the amplitude of oscillations as the viscosity of the surrounding fluid increases. 

In case of a ``free-surface'' bubble, the dynamics of the nearby free-surface as the bubble expands and collapses is studied for a wide range of {$h_{ND}$} values for each of the five liquids. A limiting stand-off distance, $h_{lim}$ is found for every liquid, for which the bubble behaves like a free-field bubble. In other words, the free-surface has a negligible effect on the bubble dynamics at a stand off distance greater than $h_{lim}$. We observe that the value of $h_{lim}$ decreases with an increase in the viscosity of the surrounding liquid and is not dependent on the bubble radius. 

For $h_{ND} < h_{lim}$, a variety of different behaviours of the free-surface such as ``spraying liquid sheet'', ``ruptured free-surface with jet on top'', ``unstable spike and unstable skirt in Deionized water'', ``unstable spike and unstable skirt in silicone oil'', ``thin spike'', ``stable spike and stable skirt'', ``spike and faster secondary jet'' and ``small rise of bulk fluid'' are observed which depends on the viscosity of the liquid and the $h_{ND}$ values. The behaviours observed from about 200 experiments are summarized in the form of a table and classified into about 8 distinct types. In general, as viscosity increases a similar type of behaviour is observed at a lower $h_{ND}$ value as compared to that for a low viscosity fluid. Some behaviours, however, are observed only at higher viscosities and not at lower viscosities. The behaviours depict a rich array of interesting sprays.

\section{\label{sec:level1}Acknowledgements}
BK and KCS thank DST, India for providing financial support through ECR/2015/000311 and MATRICS: MTR/2017/000029, respectively.

\end{document}